\documentclass[prd,nofootinbib,english]{revtex4}

\usepackage{graphicx,float}
\usepackage{amsmath,amssymb,amsfonts}
\usepackage{epsfig,color}
\usepackage[thinlines]{easytable}
\usepackage{pdfpages}
\usepackage{array}
\usepackage{cancel}
\usepackage{mathtools}
\usepackage{accents}
\usepackage{subfigure}
\usepackage{enumitem}
\usepackage[dvipsnames]{xcolor}
\usepackage{refstyle}
\usepackage{hyperref}
\usepackage{verbatim}
\hypersetup{
	colorlinks=true,
	linkcolor=blue,
	filecolor=magenta,
	citecolor=blue
}

\def\be{\begin{equation}}
\def\ee{\end{equation}}
\def\bea{\begin{eqnarray}}
\def\eea{\end{eqnarray}}

\begin{document}
\hfill  USTC-ICTS/PCFT-22-08

\title{Possible consistent model of parity violations in the symmetric teleparallel gravity}
\author{Mingzhe Li}
\email{limz@ustc.edu.cn}
\author{Yeheng Tong}
\email{yhtong19@mail.ustc.edu.cn}
\author{Dehao Zhao}
\email{dhzhao@mail.ustc.edu.cn}
\affiliation{Interdisciplinary Center for Theoretical Study, University of Science and Technology of China, Hefei, Anhui 230026, China}
\affiliation{Peng Huanwu Center for Fundamental Theory, Hefei, Anhui 230026, China}

\begin{abstract}
Many parity violating gravity models suffer from the ghost instability problem. In this paper, we consider a symmetric teleparallel gravity model that extends the general relativity equivalent model by several parity violating interactions between the gravitational field and a scalar field. These interactions exclude higher derivatives and are quadratic in the nonmetricity tensor. Through investigations on the linear cosmological perturbations, our results show that, in general, this model suffers from the difficulty caused by the existence of ghost mode in the vector perturbations. However, in cases where a special condition is imposed on the coefficients of these parity violating interactions, this model can be ghost free. 
\end{abstract}

\maketitle

\section{Introduction}

In recent years, there are many interests in investigating possible parity violations in gravity theories in the literature, partly stimulated by the experimental detections of gravitational waves (GWs) \cite{ligo1,ligo2}
and the developments in the cosmic microwave background radiation (CMB) experiments \cite{CMB1,CMB2}.
A famous and frequently studied parity violating gravity model is the so-called Chern-Simons (CS) modified gravity \cite{CSgravity1,CSgravity2},
which within the framework of Riemannian geometry modifies general relativity (GR) by a gravitational CS term
$\phi R\tilde{R}$, where $R\tilde{R}\equiv\varepsilon^{\mu\nu\rho\sigma}R_{\mu\nu}^{~~~\alpha\beta}R_{\rho\sigma\alpha\beta}$,
$\phi$ is a scalar field,
$R_{\mu\nu\rho\sigma}$ is the Riemann tensor constructed from Levi-Civita connection,
$\varepsilon^{\mu\nu\rho\sigma}=\epsilon^{\mu\nu\rho\sigma}/\sqrt{-g}$ is the Levi-Civita tensor,
$\epsilon^{\mu\nu\rho\sigma}$ is the totally antisymmetric symbol and $g$ is the determinant of the metric.
The CS modified gravity makes a difference between the amplitudes of the left- and right-handed polarized components of GWs,
but no difference between their velocities. This is the so-called amplitude birefringence phenomenon.
However, the CS modified gravity suffers from the problem of vacuum
instability because one of the circularly polarized components of GWs becomes a ghost at high frequencies \cite{CSgravity3}, i.e., its kinetic term has the wrong sign.
Further extensions to the CS modified gravity were made in Refs.~\cite{Crisostomi:2017ugk,Gao:2019liu,Zhao:2019xmm}, but these did not stop the ghost mode at high frequencies, as shown explicitly in Ref. \cite{Bartolo:2020gsh}. It seems to be difficult to have a ghost-free parity violating gravity model within the framework of Riemannian geometry.

To search for possible consistent parity violating gravity models, we may go beyond the Riemannian geometry.
Along this way, the Nieh-Yan modified Teleparallel Gravity (NYTG) model \cite{PVtele1,PVtele2} was proposed.
The NYTG model is based on the teleparallel gravity (TG) \cite{Tele,tele2021} which may be considered as a constrained metric-affine theory and is formulated in a spacetime endowed with a metric compatible but curvature free connection, the gravity is identified with the spacetime torsion. One may have a GR equivalent model within the framework of TG (we simply call it TGR because we usually use $T$-tensor to represent the torsion). The NYTG model \cite{PVtele1,PVtele2}
modifies TGR slightly by the anomalous coupling $\phi \mathcal{T}\widetilde{\mathcal{T}}$ between an axionlike field (it is a pseudo scalar field) $\phi(x)$ and the Nieh-Yan density \cite{Nieh:1981ww}: $\mathcal{T}\widetilde{\mathcal{T}}=(1/2)\varepsilon^{\mu\nu\rho\sigma}\mathcal{T}^{\lambda}_{~\mu\nu}\mathcal{T}_{\lambda\rho\sigma}$ with $\mathcal{T}^{\lambda}_{~\mu\nu}$ being the torsion tensor. The Nieh-Yan density is parity odd, so at the background with $\partial_{\mu}\phi\neq 0$, the Nieh-Yan coupling term $\phi \mathcal{T}\widetilde{\mathcal{T}}$ violates parity spontaneously.
When applying the NYTG model to cosmology \cite{PVtele1,PVtele2}, it was found that around the Friedmann-Roberson-Walker (FRW) background, there is a difference between the propagating velocities of the left- and right-handed polarized components of GWs, but the damping rates of their amplitudes are the same. This is the so-called velocity birefringence phenomenon. More importantly the NYTG model is ghostfree. Recently, the NYTG model was found to be compatible with the results of most local tests in the solar system at the post-Newtonian order \cite{Rao:2021azn,Qiao:2021fwi},
the upper limit on its model parameters by the GWs data of LIGO/Virgo Collaboration was obtained in Ref. \cite{Wu:2021ndf}, and the enhancement of primordial GWs during inflation due to the velocity birefringence of NYTG model and its implications in the air-based GWs experiments were studied in Ref. \cite{Cai:2021uup}. More recently, it was found in Ref. \cite{Li:2022mti} that generalizations of NYTG by considering the couplings of $\phi$ to other parity-odd terms quadratic in the torsion tensor \cite{Hohmann:2020dgy} are perplexed by the ghost problem again. It means in this scenario, the NYTG model is the unique one that can avoid the ghost problem. In addition, the application of Nieh-Yan term on the big-bounce cosmology was considered in Ref.~\cite{Bombacigno:2021bpk}.
Recent constraints on the parity violations in gravities from observational data can be found in Ref. \cite{Gong:2021jgg}.

Besides TG, there is another similar non-Riemannian framework to build gravity models, \textit{e.g.}, the so-called symmetric teleparallel gravity (STG) \cite{J.M.N:1999}. The STG models are formulated in spacetime endowed with a metric and a connection which is curvature free and torsionless, and gravity is identical to the nonmetricity. One may have a GR equivalent model within the framework of STG (we simply call it QGR because we usually use $Q$-tensor, \textit{e.g.}, $Q_{\alpha\mu\nu}=\hat{\nabla}_\alpha g_{\mu\nu}$,  to express the nonmetricity). Similar to the CS modified gravity and the NYTG model, the simplest parity violating extension to QGR is given by the coupling: $\phi(x)\varepsilon^{\mu\nu\rho\sigma}Q_{\mu\nu\alpha}Q_{\rho\sigma}^{~~~\alpha}$. This modification indeed produces velocity birefringence phenomenon in the cosmological tensor perturbations \cite{Conroy:2019ibo}, but it was found in Ref. \cite{ STGPV2} that it suffers from severe theoretical problems when cosmological vector perturbations are considered. As was pointed out in Ref. \cite{ STGPV2}, this parity violating extension promotes the vector perturbations to dynamical degrees of freedom and also causes velocity and amplitude birefringence phenomena in the vector modes, and more importantly, one of the vector modes becomes ghost at high momentum scales. It seems that the amplitude birefringence is always accompanied by the ghost instability. 

In this paper, within the STG framework we consider a more general model which extends QGR by several parity violating interactions between the scalar field and the parity-odd terms, which are quadratic in the nonmetricity tensor, \textit{i.e.} $\mathcal{L}\sim\varepsilon QQ\hat{\nabla}\phi\hat{\nabla}\phi$. These interactions have been considered in \cite{Conroy:2019ibo}, but only the tensor perturbations were studied there. Though no dangerous mode was found in the tensor perturbations, this cannot guarantee the absence of pathologies in the scalar and vector perturbations. In this paper, we will make detailed investigations on the linear cosmological perturbations of this model, including all the scalar, vector, and tensor types of perturbations. We will attach much importance to the vector perturbations and show in what conditions this model can be free from the ghost problem.

This paper is organized as follows. In Sec. \ref{section basic knowledge}, we will introduce the STG model with parity violations we will consider in this paper. In Sec. \ref{cosmology}, we will apply this model to cosmology and present our main results about the studies on the linear cosmological perturbations. Section \ref{conclusion} is a summary.

\section{The Symmetric Teleparallel Gravity Model with Parity Violations}\label{section basic knowledge}

We will take the unit $8\pi G=1$ and the convention for the metric signature: $\left\lbrace +,-,-,-\right\rbrace$. As usual, the Greeks $\mu,\nu,\rho,\dots=0,1,2,3$ are used to represent spacetime tensor indices, and the Latins $i,j,k,\dots=1,2,3$ are used to denote the spatial components. As mentioned before, the STG theory is considered as a constrained metric-affine theory, it is formulated by a metric $g_{\mu\nu}$ and an affine connection ${\Gamma^\lambda}_{\mu\nu}$, which is curvature free and torsionless,
\begin{equation}
	{\hat{R}^\rho}_{\sigma\mu\nu}\equiv\partial_\mu{\Gamma^\rho}_{\nu\sigma}+\Gamma^\rho_{~\mu\alpha}{\Gamma^\alpha}_{\nu\sigma}-\left\lbrace\mu\leftrightarrow\nu\right\rbrace=0~,~\mathcal{T}^\rho_{~\mu\nu}\equiv\Gamma^\rho_{~\mu\nu}-\Gamma^\rho_{~\nu\mu}=0.
\end{equation}
With these constraints, the affine connection can be generally expressed as 
\begin{equation}
	{\Gamma^\lambda}_{\mu\nu}=\frac{\partial x^\lambda}{\partial y^\alpha}\partial_\mu\partial_\nu y^\alpha~,
\end{equation}
where the four functions $y^\alpha$ themselves form a special coordinate system in which all the components of the affine connection vanish. One can carry out the calculations by fixing to this special coordinate system. In fact, such  "coincident gauge" had been frequently adopted in the literature. However, for the purpose of making our analysis as general as possible, we prefer to work in an arbitrary coordinate system in this paper. We will consider the metric $g_{\mu\nu}$ and the four functions $y^\mu$ as the fundamental variables of the STG theory.

Within STG, the gravity is attributed to the nonmetricity tensor $Q_{\alpha\mu\nu}\equiv\hat{\nabla}_\alpha g_{\mu\nu}=\partial_\alpha g_{\mu\nu}-{\Gamma^\lambda}_{\alpha\mu}g_{\lambda\nu}-{\Gamma^\lambda}_{\alpha\nu}g_{\mu\lambda}$, 
which measures the failure of the connection to be metric compatible. The QGR model, which is equivalent to GR within the STG framework, has the following action:
\begin{equation}\label{Qac}
	S_g=\frac{1}{2}\int d^4x \sqrt{-g}\mathbb{Q} \equiv \frac{1}{2}\int d^4x \sqrt{-g} \left( \frac{1}{4}Q_{\alpha\mu\nu}Q^{\alpha\mu\nu}-\frac{1}{2}Q_{\alpha\mu\nu}Q^{\mu\nu\alpha}-\frac{1}{4}Q_{\alpha}Q^{\alpha}+\frac{1}{2}Q_{\alpha}\bar{Q}^{\alpha}\right)~,
\end{equation}
where $Q_\alpha=Q_{\alpha\mu\nu}g^{\mu\nu}$ and $\bar{Q}_\alpha=Q_{\rho\sigma\alpha}g^{\rho\sigma}$ are two nonmetricity vectors. This action is identical to the Einstein-Hilbert action up to a boundary term, 
\begin{equation}
	S_g=\frac{1}{2}\int d^4x \sqrt{-g} \left[-R-\nabla_{\alpha}(Q^\alpha-\bar{Q}^\alpha)\right],
\end{equation}
where both the curvature scalar $R$ and the covariant derivative $\nabla_{\alpha}$ is associated with the Levi-Civita connection. 

Similar to the CS modified gravity and the NYTG model,  a simple parity violating extension to QGR can be realized by introducing an extra term 
$S_{PV}\sim \int d^4x\sqrt{-g} \phi\varepsilon^{\mu\nu\rho\sigma}\,Q_{\mu\nu\alpha}\,{Q_{\rho\sigma}}^\alpha$ to the action (\ref{Qac}). 
However, as pointed out in Ref. \cite{ STGPV2}, with this modification, the vector perturbations in the gravity sector are promoted to be propagating dynamical modes, and one of the vector modes becomes a ghost at high momentum scales.

Now we turn to the more general (and of course more complex) case where parity violating extensions in gravity are provided by several interactions between the scalar field and the parity-odd terms which are quadratic in the nonmetricity tensor. At the same time, we should guarantee the equations of motion of the metric and the scalar field to be second order. This is in order to exclude the Ostrogradsky ghosts, which are originated from higher derivatives. Totally there are seven couplings satisfying the requirements, as listed below,
\begin{equation}
	\begin{aligned}\label{PVc}
		M_1&=\varepsilon^{\mu\nu\rho\sigma}\,Q_{\mu\nu\alpha}\,{Q_{\rho\sigma}}^\alpha\,\nabla_\beta \phi\,\nabla^\beta \phi,\\
		M_2&=\varepsilon^{\mu\nu\rho\sigma}\,Q_{\mu\nu\alpha}\,{Q_{\rho\sigma}}^\beta\,\nabla^\alpha \phi\,\nabla_\beta \phi,\\
		M_3&=\varepsilon^{\mu\nu\rho\sigma}\,Q_{\mu\nu\alpha}\,{Q_{\rho}}^{\alpha\beta}\,\nabla_\sigma \phi\,\nabla_\beta \phi,\\
		M_4&=\varepsilon^{\mu\nu\rho\sigma}\,Q_{\mu\nu\alpha}\,{Q^{\alpha\beta}}_\rho\,\nabla_\sigma \phi\,\nabla_\beta \phi,\\
		M_5&=\varepsilon^{\mu\nu\rho\sigma}\,Q_{\mu\nu\alpha}\,{Q^{\beta\alpha}}_\rho\,\nabla_\sigma \phi\,\nabla_\beta \phi,\\
		M_6&=\varepsilon^{\mu\nu\rho\sigma}\,Q_{\mu\nu\alpha}\,Q_{\rho}\,\nabla^\alpha\phi\,\nabla_\sigma \phi,\\
		M_7&=\varepsilon^{\mu\nu\rho\sigma}\,Q_{\mu\nu\alpha}\,\bar{Q}_{\rho}\,\nabla^\alpha\phi\,\nabla_\sigma \phi.
	\end{aligned}
\end{equation}
Then the action of parity violation is written as
\begin{equation}\label{PVa}
	S_{PV}=\int d^4x\sqrt{-g}\sum_{a} c_a(\phi,\nabla^\mu \phi \nabla_\mu \phi) M_a\equiv\sum_{a} S_{PVa}~,
\end{equation}
with $a=1,2,\cdots,7$. The coupling coefficients $c_a$ can also rely on the scalar field and its first derivatives.
So the full action of the model under consideration is 
\begin{equation}
	S=S_g+S_{PV}+S_\phi= \int d^4x\sqrt{-g}\left[\frac{\mathbb{Q}}{2}+\sum_{a} c_a(\phi,\nabla^\mu \phi \nabla_\mu \phi) M_a +\frac{1}{2} g^{\mu\nu}\partial_\mu\phi\partial_\nu\phi-V(\phi)\right] \label{S2}~,
\end{equation}
where we have neglected the matter other than the scalar field $\phi$ in the Universe. The model studied in Ref. \cite{ STGPV2} is equivalent to a special case with $c_2=c_3=...=c_7=0$.
This more general model was considered in \cite{Conroy:2019ibo}, where only the tensor modes of the linear cosmological perturbation were considered. However, as a modified gravity model, only considering tensor perturbations is not enough. In the following, we will make a full investigation on its linear cosmological perturbations around the FRW universe by considering all the scalar, vector and tensor perturbations. 

\section{Application to Cosmology}\label{cosmology}
\subsection{The background dynamics}

For simplicity, we take the spatially flat FRW universe as the background, its line element is given by 
\begin{equation}
	ds^2=a^2(\eta)(d\eta^2-\delta_{ij}dx^i dx^j),
\end{equation}
where $\eta$ is the conformal time. In GR we only need the metric to define a background, since only the metric is the fundamental variable; while in the STG theory, the four functions $y^\mu$ determining affine connection are also considered as fundamental variables. Hence, besides the metric, these four functions should also be given. According to \cite{2104.02483}, $\bar{\Gamma}^{\lambda}_{~\mu\nu}=0$ is a good solution (not a gauge choice) to the background evolution equation, which means we can set $y^\mu =x^\mu\equiv\left\lbrace\eta,x^i \right\rbrace$ for the cosmic background. There are other solutions for the affine connection compatible with the metric ansatz in above line element, for instances, the nontrivial solutions found in Refs. \cite{Hohmann:2021ast, DAmbrosio:2021pnd}, but in this paper we only consider the simplest one. After some calculations, one can straight forwardly find that the extra parity violating terms in the model (\ref{S2}) do not have effects on the evolution of the background. Thus the background dynamics is the same as in GR, 
\begin{equation}
	3\mathcal{H}^2=a^2\rho_\phi,\quad-2\mathcal{H}'-\mathcal{H}^2=a^2 p_\phi,\quad\phi''+2\mathcal{H}\phi'+a^2\phi=0,
\end{equation}
where the prime represents the derivative with respect to the conformal time $\eta$ and $\mathcal{H}=a'/a$ is the conformal Hubble rate. As usual, $\rho_\phi= \phi'^2/(2a^2)+V(\phi)$ and $p_\phi=\phi'^2/(2a^2)-V(\phi)$ are the energy density and pressure of the scalar field, respectively.

\subsection{Cosmological perturbations}

Now we consider the linear cosmological perturbations of the model (\ref{S2}). With the Scalar-Vector-Tensor decomposition, the perturbed metric is contained in the following parametrized line element:
\begin{equation}
	ds^2=a^2\left\lbrace (1+2A)d\eta^2+2(\partial_i B+B_i)d\eta dx^i-\left[(1-2\psi)\delta_{ij}+2\partial_i\partial_jE+\partial_jE_i+\partial_iE_j+h_{ij}\right]dx^i dx^j \right\rbrace~,
\end{equation}
where $A,\psi,B,E$ are the scalar perturbations, $E_i, B_i$ are the vector perturbations which satisfy the transverse conditions: $\partial_i B_i=\partial_i E_i=0$, and the tensor perturbations $h_{ij}$ are transverse and traceless: $\partial_i h_{ij}=0$ and $\delta^{ij}h_{ij}=0$.

Since the connection vanishes on the background, itself is considered as the perturbation. The four functions used to construct the connection can be decomposed as $y^\alpha=x^\alpha+u^\alpha$, where the perturbation $u^\alpha$ is further decomposed as  $u^\alpha\equiv\left\lbrace u^0,\partial_i u+u_i \right\rbrace$ with the transverse condition: $\partial_i u_i=0$. So we have two more scalar perturbations $u^0$ and $u$, and one more vector perturbation $u_i$. Then, up to the second order the perturbed connection built from $y^\alpha$ is
\begin{equation}
	{\Gamma^\lambda}_{\mu\nu}=\frac{\partial x^\lambda}{\partial y^\alpha}\partial_\mu \partial_\nu y^\alpha =\partial_\mu \partial_\nu u^\lambda -\partial_\mu \partial_\nu u^\alpha \partial_\alpha u^\lambda.
\end{equation}
As usual, the background and perturbation decomposition for the scalar field is: $\phi(\eta,\vec{x})=\phi(\eta)+\delta\phi(\eta,\vec{x})$.

One can get the equations of motion by varying the action (\ref{S2}) with respect to the metric $g_{\mu\nu}$, the functions $y^{\alpha}$ and the scalar field $\phi$, then using the background-perturbations decomposition to obtain the equations for the scalar, vector and tensor perturbations, respectively. For our purpose in this paper, it is better to consider directly the quadratic actions for the perturbations. The linear perturbation equations can be obtained from the quadratic actions through the variational principle, so all the properties of the perturbation equations have been contained in the quadratic actions already. Furthermore, the quadratic action has an advantage of showing clearly whether there are dangerous modes in the spectrum. Hence, in the following subsections, we will focus on the quadratic actions for the perturbations of the model (\ref{S2}).

\subsection{Quadratic actions for scalar and tensor Perturbations}

For the scalar perturbations, it is not difficult to find that all the parity violating terms in Eq. (\ref{PVc}) vanish up to the second order. So, they are at least third order quantities and have no contribution to the quadratic action. Therefore, the quadratic action for the scalar perturbations of the model (\ref{S2}) is the same as the one in GR with a minimally coupled scalar field:
\begin{equation}
	S^{(2)}_S= \int d^4x \frac{a^2\phi'^2}{2\mathcal{H}^2}\left(\zeta'^2-\partial_i\zeta\partial_i\zeta\right)~,
\end{equation}
where $\zeta=-(\psi+\mathcal{H}\delta\phi/\phi')$ is a gauge invariant variable and denotes the curvature perturbation of the hypersurface with the constant $\phi$ field. 

For tensor perturbations, one can find that only the first and the fifth parity violating terms of Eq. (\ref{PVc}) have non-zero contributions to the quadratic action:  
\begin{equation}
	S_{PV1}=-\int d^4x\,2c_1\,\phi'^2 \epsilon_{ijk}h_{jl,i}h'_{kl}~,~ S_{PV5}=\int d^4x\,c_5\,\phi'^2 \epsilon_{ijk}h_{jl,i}h'_{kl}~,
\end{equation}
where $\epsilon_{ijk}$ is the 3-dimensional anti-symmetric symbol and $\epsilon_{123}=-1$. So the full quadratic action for tensor perturbations of the model (\ref{S2}) is
\begin{equation}\label{qSt}
	S^{(2)}_T= \int d^4x \left\lbrace \frac{a^2}{8}\left( h'_{ij}h'_{ij} -h_{ij,k} h_{ij,k}\right) +\frac{1}{2}[(2c_1-c_5)\phi'^2]' \epsilon_{ijk}h_{jl,i}h_{kl}\right\rbrace~.
\end{equation}
The parity violations can be shown more clearly in the Fourier space. For this, we expand the tensor perturbations $h_{ij}$ in terms of the plane wave and the circular polarization bases $e^A_{ij}$,
\begin{equation}
	h_{ij}=\sum_{A=L,R}\int d\eta d^3\vec{k}\,e^{-i\vec{k}\cdot\vec{x}}\,h^A(\eta,\vec{k})\,e^A_{ij}~,
\end{equation}
the index $A=L,R$ denotes the left- and right-handed polarizations, the bases $e^A_{ij}$ satisfy the relations $e^A_{ij}e^{B*}_{ij}=\delta^{AB}$, and $i\epsilon_{ijk}k_i e^A_{jl}=\lambda_A k e^A_{kl}$. 
The parameter $\lambda_A=\mp 1$ for $A=L, R$, respectively, and is used to remind the helicity dependence. Then the quadratic action (\ref{qSt}) is rewritten as
\begin{equation}\label{qSt1}
	S^{(2)}_T=\sum_{A=L,R} \int d\eta d^3\vec{k} \frac{a^2}{8} (h'^A h'^{A*}-\omega^{2}_{AT}h^A h^{A*})~,
\end{equation}
with
\begin{equation}
	\omega^2_{AT}=k^2\left\lbrace 1+\frac{4\lambda_A}{a^2k}[(2c_1-c_5)\phi'^2]'\right\rbrace \label{drT}~.
\end{equation}
The quadratic action (\ref{qSt1}) showed that there is no ghost mode in the tensor perturbations because both the left- and right-handed polarization modes of GWs have the right sign in their kinetic terms. But the dispersion relation (\ref{drT}) is helicity dependent. This implies a velocity difference between the two circular polarization modes, i.e., the velocity birefringence phenomenon.  
Both polarization modes have propagating velocities different from the speed of light, so are constrained by the event GW170817 observed by LIGO/Virgo \cite{ligo2}; this in turn puts constraints on the coefficients $c_1$ and $c_5$ of this model \cite{Conroy:2019ibo}. 

\subsection{Quadratic action for vector perturbations}

We have seen that the scalar perturbations of the model (\ref{S2}) are trivial, and its tensor perturbations have been considered in \cite{Conroy:2019ibo}. The vector perturbations of this model have not been explored, and their properties are our main interests in this paper. 

After some tedious calculations, we find the following five parity violating terms have contributions to the quadratic action for the vector perturbations: 
\begin{equation}
\begin{aligned}
	S_{PV1}&=\int d^4x 2c_1 \phi'^2 \epsilon_{ijk}\left[ (B_{j,i}+u'_{j,i})(B'_k+u''_k)-(E_{j,il}-u_{j,il})(E'_{k,l}-u'_{k,l}) \right],\\
	S_{PV2}&=\int d^4x 2c_2 \phi'^2 \epsilon_{ijk}(B_{j,i}+u'_{j,i})(B'_k+u''_k), \\
	S_{PV4}&=-\int d^4x c_4 \phi'^2 \epsilon_{ijk}\left[ (B_{j,i}+u'_{j,i})(B'_k+u''_k) +(E_{j,il}-u_{j,il})(B_{k,l}+u'_{k,l})\right], \\
	S_{PV5}&=\int d^4x c_5 \phi'^2 \epsilon_{ijk}\left[(E_{j,il}-u_{j,il})(E'_{k,l}-u'_{k,l}) -(B_{j,i}+u'_{j,i})(B'_k+u''_k)\right], \\
	S_{PV7}&=-\int d^4x c_7 \phi'^2 \epsilon_{ijk}(B_{j,i}+u'_{j,i})\left[(B'_k+u''_k) +\nabla^2(E_k-u_k)\right],
\end{aligned}
\end{equation}
where $\nabla^2=\partial_i\partial_i$ is the Laplacian. In order to simplify the calculations, we will fix a gauge before further investigations. In this subsection, we will take the coincident gauge: $\Gamma^{\mu}_{~\rho\nu}=0$, this means the four functions $y^{\mu}$ used to construct the affine connection match the coordinates $x^{\mu}$, i.e., $y^{\mu}=x^{\mu}$, to all orders. So the vector perturbations $u_i$ separated from $y^{\mu}$ vanish identically, $u_i=0$. After some calculations under the coincident gauge, we have the full quadratic action for the vector perturbations of the model (\ref{S2}), 
\begin{equation}
	S^{(2)}_V=\int d^4x \left[ \frac{1}{4}a^2(B_{i,j}+E'_{i,j})(B_{i,j}+E'_{i,j}) +\epsilon_{ijk}\phi'^2\left( b_1 B'_k B_{j,i} -b_2 B_{k,l} E_{j,il} -b_3 E'_{k,l} E_{j,il}\right) \right]\label{qSv}~,
\end{equation}
where we have defined $b_1=2c_1+2c_2-c_4-c_5-c_7$, $b_2=c_4-c_7$ and $b_3=2c_1-c_5$.

It is clear from above quadratic action that the variables $B_i$ are not dynamical modes, the variation of (\ref{qSv}) with respect to $B_i$ yields the following constraint equation:
\begin{equation}
	-\frac{1}{2}a^2\nabla^2(B_k+E'_k) -\epsilon_{ijk}\left(b_1\phi'^2\right)'B_{j,i} +\epsilon_{ijk}b_2 \phi'^2 \nabla^2 E_{j,i}=0.
\end{equation}
It is better to solve this equation in the Fourier space. Similar to what we have done in the previous subsection for the tensor perturbations, one can expand the perturbations $B_i$ and $E_i$ in terms of the plane wave and the circular polarization bases $e^A_i$ with $A=L,R$ as
\begin{equation}\label{constraint}
	B_i\equiv\int d\eta d^3\vec{k}\,e^{-i\vec{k}\cdot\vec{x}}\,B^A(\eta,\vec{k})\,e^A_i~,
\end{equation}
and a same expansion for $E_i$. Similarly, the circular polarization bases for vector perturbations $e^A_i$ satisfy the relations $e^A_i e^{B*}_i=\delta^{AB}$ and $ik_i e^A_j \epsilon_{ijk}=k\lambda^A e^A_k$. Again, the parameters $\lambda^A=\mp 1$ for $A=L, R$ remind the helicity dependence when they appear.  
With these equipments, the constraint equation (\ref{constraint}) is rewritten in the following form:
\begin{equation}
	a^2k^2 (B^A+E'^A)+2k\lambda^A\left[\left(b_1\phi'^2\right)'B^A +b_2 \phi'^2 k^2 E^A\right]=0~,
\end{equation}
and can be solved as
\begin{equation}\label{BA1}
	B^A= -\frac{2\lambda^A b_2 \phi'^2 k^2}{a^2 k+2\lambda^A \left(b_1\phi'^2\right)'}E^A-\frac{a^2 k}{a^2 k+2\lambda^A \left(b_1\phi'^2\right)'} E'^A~.
\end{equation}
At the same time, we also rewrite the quadratic action (\ref{qSv}) in the Fourier space:
\begin{equation}
\begin{aligned}
	S^{(2)}_V&= \sum_{A=L,R} \int d\eta d^3\vec{k} \left\lbrace \frac{1}{4} a^2 k^2(B^A+E'^A)(B^{*A}+E'^{*A})+\frac{1}{2}\lambda^A k\left[ \left(b_1\phi'^2\right)'B^A B^{*A}\right.\right.\\
	&\left.\left.\quad\quad\quad\quad\quad\quad\quad\quad +2b_2\phi'^2 k^2 E^A B^{*A}-\left(b_3\phi'^2\right)'k^2 E^A E^{*A}\right] \right\rbrace.
\end{aligned}
\end{equation}
Then substitute the solution (\ref{BA1}) back into the action (\ref{qSv}), one finally obtains the quadratic action for the vector perturbations, 
\begin{equation}\label{BV1}
	S^{(2)}_V= \sum_{A=L,R} \int d\eta d^3k \, \left( z_A^2 E'^A E'^{*A} -w_A^2 E^A E^{*A}\right)~,
\end{equation}
where
\begin{equation}\label{kc}
\begin{aligned}
	z_A^2&= \frac{\lambda^A \left(b_1\phi'^2\right)' a^2 k^2}{2a^2 k+4\lambda^A \left(b_1\phi'^2\right)'}~,\\ w_A^2&= \frac{a^2 k^4 b_2 \phi'^2 \left(b_1 \phi'^2\right)''}{\left[ a^2 k+2\lambda^A \left(b_1 \phi'^2\right)'\right]^2} +\frac{2k^5 b_2^2 \phi'^4 -\lambda^A a^2 k^4 \left(b_2 \phi'^2\right)'}{2a^2 k+4\lambda^A \left(b_1 \phi'^2\right)'} +\frac{1}{2}\lambda^A k^3 \left(b_3 \phi'^2\right)'~.
\end{aligned}
\end{equation}

The key properties can be read out from the action (\ref{BV1}) and Eq. (\ref{kc}). First, as long as $(b_1\phi'^2)'\neq 0$, the factors $z_A^2$ with $A=L, R$ in front of the kinetic terms of $E^A$ do not vanish, so both components of the vector perturbations represent dynamical degrees of freedom. They propagate in the Universe since generating in the very early time (e.g., inflation epoch). This is different from the model of GR with a coupled scalar field, where both components of vector perturbations just represent some constraints. Second, $z_A^2$ depend on the helicity as well as the wave number $k$. The former dependence induces different damping rates for different polarization modes when they propagating in the Universe and implies the amplitude birefringence phenomenon. The dependence on $k$ means the behaviors of the polarization modes are scale dependent. From Eq. (\ref{kc}), one can see that at large length scales, where $k\ll 2a^{-2}\left| \left(b_1\phi'^2\right)'\right| $, the factors $z_A^2\simeq a^2 k^2/4>0$ are positive and helicity independent. So, both polarization modes are healthy. However, at small length scales, where $k\gg 2 a^{-2}\left| \left(b_1\phi'^2\right)'\right| $, $z_A^2\simeq\lambda^A \left(b_1\phi'^2\right)' k/2$.  We know $\lambda^A=\pm 1$, as long as $\left(b_1\phi'^2\right)'\neq0$, one of the two components must be a ghost mode because it receives a minus sign in its kinetic term. Third, one may define $\omega_{A}^2\equiv w_A^2/z_A^2$, their dependences on the wave number $k$ define the dispersion relations. In general, the dispersion relations are expected to be helicity dependent; this implies the velocity birefringence phenomenon. In all, as long as $(b_1\phi'^2)'\neq 0$, the vector perturbations in this model (\ref{S2}) are dynamical, present both amplitude and velocity birefringence phenomena. The worse consequence is that one of the circularly polarized components of the vector perturbations is a ghost mode and leads to vacuum instability. All these results have been shown in \cite{STGPV2} for a simpler version of this model. 

The above discussions at the same time pointed out the way circumventing these difficulties (at least for linear perturbation theory): that is $z_A^2=0$ or $\left(b_1\phi'^2\right)'=0$, so that $b_1= {\rm const.}/\phi'^2$. To have a nonzero $b_1$ satisfying such requirement will need to fine tune the scalar field model. A simpler choice is to have a vanished $b_1$, i.e., $b_1=2c_1+2c_2-c_4-c_5-c_7=0$. Under this condition, neither components of vector perturbations are dynamical fields; they just represent constraints, as same as the model of GR with a coupled scalar field. In this case, the problem of ghost instability certainly does not exist. 
Through the same procedure, one may find that in the cases of $b_1=0$, 
the quadratic action (\ref{qSv}) reduces to
\begin{equation}
	S^{(2)}_V=- \sum_{A=L,R} \int d\eta d^3k w_A^2 E^A E^{*A}~,
\end{equation}
where
\begin{equation}
	w_A^2=\frac{k^4}{a^2} b_2^2 \phi'^4 +\frac{1}{2}\lambda^A k^3 \left[(b_3-b_2)\phi'^2\right]'~.
\end{equation}
It is evidently an action for non-dynamical fields.

\subsection{Another approach: conformal-Newtonian gauge}

In the literature, many works on the cosmological perturbations of the STG models adopted the coincident gauge to simplify the calculations, just like we have done in the previous subsection. In this subsection we take another gauge used extensively in GR, \textit{i.e.} the conformal-Newtonian gauge, under which $E_i=0$. We expect these two approaches should give the same results.

Under the conformal-Newtonian gauge, the quadratic action for the vector perturbations of the model (\ref{S2}) is
\begin{equation}
	S^{(2)}_V=\int d^4x \left\lbrace \frac{1}{4}a^2 B_{i,j} B_{i,j}+\epsilon_{ijk}\phi'^2\left[ b_1(B'_k+u''_k)(B_{j,i}+u'_{j,i}) +b_2 B_{k,l} u_{j,il} -(b_3-b_2) u'_{k,l} u_{j,il}\right] \right\rbrace\label{qSv2}
\end{equation}
where $b_1$, $b_2$ and $b_3$ are defined as same as before. With similar operations in the previous subsection, the variation of (\ref{qSv2}) with respect to $B_k$ gives
\begin{equation}
	-\frac{1}{2}a^2\nabla^2 B_k -\epsilon_{ijk}\left[\left(b_1\phi'^2\right)'(B_{j,i}+u'_{j,i}) +b_2 \phi'^2 \nabla^2 u_{j,i}\right] =0
\end{equation}
After transforming this constraint equation to the Fourier space, one can solve $B^A$ as
\begin{equation}\label{BA2}
	B^A= 2\lambda^A\frac{-\left(b_1\phi'^2\right)'u'^A +b_2\phi'^2 k^2 u^A}{a^2 k+2\lambda^A \left(b_1\phi'^2\right)'}
\end{equation}
Then substitute (\ref{BA2}) back into the action (\ref{qSv2}), one obtains the final result for the quadratic action, 
\begin{equation}\label{BV2}
	S^{(2)}_V= \sum_{A=L,R} \int d\eta d^3k \left[z_A^2 u'^A u'^{*A} -w_A^2 u^A u^{*A}\right]~,
\end{equation}
where $z_A^2$ and $w^2_{A}$ are the same as those in Eq. (\ref{kc}) in the last subsection. 

So, for the vector perturbations we can define gauge invariant variables $V_i=E_i-u_i$, or $V^A=E^A-u^A$ under the circular polarization bases. Thus, in the conformal-Newtonian gauge, $V_i=-u_i$, and in the coincident gauge, $V_i=E_i$. In practice, if we do not take any gauge during the calculations, we finally get a quadratic action for the gauge invariant vector perturbations. This action has the same form as those of the actions (\ref{BV1}) and (\ref{BV2}), except replacing $E^A$ in Eq.  (\ref{BV1}) and $u^A$ in Eq.  (\ref{BV2}) by the variable $V^A$. 

\section{Conclusion}\label{conclusion}

In this paper, we studied a STG model (\ref{S2}) which modifies QGR (\ref{Qac}) by several parity violating interactions between the gravitational field and a scalar field. These extra interactions are quadratic in the nonmetricity tensor and do not introduce any higher derivatives. Through applying this model to cosmology, we found that these modifications do not change the background dynamics and the scalar perturbations. They brought changes to the tensor perturbations and caused velocity birefringence in GWs. We attached much importance on the vector perturbations of this model. Our results showed that, in general, the vector perturbations are promoted to be dynamical propagating fields and present both amplitude and velocity birefringence phenomena. The worst consequence is that one of the circularly polarized components of the vector perturbations becomes a ghost at small length (large momentum) scales. All these features also appeared in a simpler version of this model \cite{STGPV2}. 

We also showed a way out of these difficulties, at least for linear perturbation theory. We found that under a special combination of the PV interaction terms, the vector perturbations do not propagate and go back to be constraints, so the model can be free from ghost modes.  In more detail, the condition to avoid ghost requires the coefficients in the action (\ref{S2}) satisfy $b_{1}\equiv 2c_1+2c_2-c_4-c_5-c_7=0$. 

We also notice that the ghostfree condition covers five of the parity violating terms only, while the rest two terms vanish up to the second order for the FRW background and have no contribution to the quadratic actions. We suspect that these two terms may have effects for linear perturbations around other backgrounds.

\section{Acknowledgments}
This work is supported in part by NSFC under Grant No. 12075231 and No. 12047502, and by National Key Research and Development Program of China Grant No. 2021YFC2203100.


\begin{thebibliography}{}

\bibitem{ligo1}
B.~P.~Abbott \textit{et al.} (LIGO Scientific and Virgo Collaborations),
Phys. Rev. Lett. \textbf{116}, no.6, 061102 (2016).
doi:10.1103/PhysRevLett.116.061102

\bibitem{ligo2}
B.~P.~Abbott \textit{et al.} (LIGO Scientific and Virgo Collaborations),
Phys. Rev. Lett. \textbf{119}, no.16, 161101 (2017).
doi:10.1103/PhysRevLett.119.161101

\bibitem{CMB1}
H.~Li, S.~Y.~Li, Y.~Liu, Y.~P.~Li, Y.~Cai, M.~Li, G.~B.~Zhao, C.~Z.~Liu, Z.~W.~Li and H.~Xu, \textit{et al.}
Natl. Sci. Rev. \textbf{6}, no.1, 145-154 (2019).
doi:10.1093/nsr/nwy019

\bibitem{CMB2}
K.~Abazajian \textit{et al.} (CMB-S4 Collaboration),
Astrophys. J. \textbf{926}, 54 (2022).
doi:10.3847/1538-4357/ac1596

\bibitem{CSgravity1}
R.~Jackiw and S.~Y.~Pi,
Phys. Rev. D \textbf{68}, 104012 (2003).
doi:10.1103/PhysRevD.68.104012

\bibitem{CSgravity2}
S.~Alexander and N.~Yunes,
Phys. Rept. \textbf{480}, 1-55 (2009).
doi:10.1016/j.physrep.2009.07.002

\bibitem{CSgravity3}
S.~Dyda, E.~E.~Flanagan and M.~Kamionkowski,
Phys. Rev. D \textbf{86}, 124031 (2012).
doi:10.1103/PhysRevD.86.124031

\bibitem{Crisostomi:2017ugk}
M.~Crisostomi, K.~Noui, C.~Charmousis and D.~Langlois,
Phys. Rev. D \textbf{97}, no.4, 044034 (2018).
doi:10.1103/PhysRevD.97.044034

\bibitem{Gao:2019liu}
X.~Gao and X.~Y.~Hong,
Phys. Rev. D \textbf{101}, no.6, 064057 (2020).
doi:10.1103/PhysRevD.101.064057

\bibitem{Zhao:2019xmm}
W.~Zhao, T.~Zhu, J.~Qiao and A.~Wang,
Phys. Rev. D \textbf{101}, no.2, 024002 (2020).
doi:10.1103/PhysRevD.101.024002

\bibitem{Bartolo:2020gsh}
N.~Bartolo, L.~Caloni, G.~Orlando and A.~Ricciardone,
J. Cosmol. Astropart. Phys. \textbf{03} (2021) 073.
doi:10.1088/1475-7516/2021/03/073

\bibitem{PVtele1}
M.~Li, H.~Rao and D.~Zhao,
J. Cosmol. Astropart. Phys. \textbf{11} (2020) 023.
doi:10.1088/1475-7516/2020/11/023

\bibitem{PVtele2}
M.~Li, H.~Rao and Y.~Tong,
Phys. Rev. D \textbf{104}, no.8, 084077 (2021).
doi:10.1103/PhysRevD.104.084077

\bibitem{Tele}
R. ~Aldrovandi and J.~G.~Pereira, {\it Teleparallel Gravity}, Vol. 173. Springer, 23 Dordrecht, (2013).

\bibitem{tele2021}
S.~Bahamonde, K.~F.~Dialektopoulos, C.~Escamilla-Rivera, G.~Farrugia, V.~Gakis, M.~Hendry, M.~Hohmann, J.~L.~Said, J.~Mifsud and E.~Di Valentino,
[arXiv:2106.13793 [gr-qc]].

\bibitem{Nieh:1981ww}
H.~T.~Nieh and M.~L.~Yan,
J. Math. Phys. \textbf{23}, 373 (1982).
doi:10.1063/1.525379

\bibitem{Rao:2021azn}
H.~Rao,
Phys. Rev. D \textbf{104}, no.12, 124084 (2021).
doi:10.1103/PhysRevD.104.124084

\bibitem{Qiao:2021fwi}
J.~Qiao, T.~Zhu, G.~Li and W.~Zhao,
[arXiv:2110.09033 [gr-qc]].

\bibitem{Wu:2021ndf}
Q.~Wu, T.~Zhu, R.~Niu, W.~Zhao and A.~Wang,
Phys. Rev. D \textbf{105}, 024035 (2022).
doi:10.1103/PhysRevD.105.024035

\bibitem{Cai:2021uup}
R.~G.~Cai, C.~Fu and W.~W.~Yu,
[arXiv:2112.04794 [astro-ph.CO]].

\bibitem{Li:2022mti}
M.~Li, Z.~Li and H.~Rao,
[arXiv:2201.02357 [gr-qc]].

\bibitem{Hohmann:2020dgy}
M.~Hohmann and C.~Pfeifer,
Eur. Phys. J. C \textbf{81}, no.4, 376 (2021).
doi:10.1140/epjc/s10052-021-09165-x

\bibitem{Bombacigno:2021bpk}
F.~Bombacigno, S.~Boudet, G.~J.~Olmo and G.~Montani,
Phys. Rev. D \textbf{103}, no.12, 124031 (2021).
doi:10.1103/PhysRevD.103.124031

\bibitem{Gong:2021jgg}
C.~Gong, T.~Zhu, R.~Niu, Q.~Wu, J.~L.~Cui, X.~Zhang, W.~Zhao and A.~Wang,
Phys. Rev. D \textbf{105}, 044034 (2022).
doi:10.1103/PhysRevD.105.044034

\bibitem{J.M.N:1999}
J.~M.~Nester and H.~J.~Yo,
Chin. J. Phys. \textbf{37}, 113 (1999).


\bibitem{Conroy:2019ibo}
A.~Conroy and T.~Koivisto,
J. Cosmol. Astropart. Phys. \textbf{12} (2019) 016.
doi:10.1088/1475-7516/2019/12/016

\bibitem{STGPV2}
M.~Li and D.~Zhao,
Phys. Lett. B \textbf{827}, 136968 (2022).
doi:10.1016/j.physletb.2022.136968
	
\bibitem{2104.02483}
D.~Zhao,
Eur. Phys. J. C \textbf{82}, 303 (2022).
doi:10.1140/epjc/s10052-022-10266-4

\bibitem{Hohmann:2021ast}
M.~Hohmann,
Phys. Rev. D \textbf{104}, no.12, 124077 (2021).
doi:10.1103/PhysRevD.104.124077
	
\bibitem{DAmbrosio:2021pnd}
F.~D'Ambrosio, L.~Heisenberg and S.~Kuhn,
Class. Quant. Grav. \textbf{39}, no.2, 025013 (2022).
doi:10.1088/1361-6382/ac3f99


	
\end{thebibliography}
\end{document}